\documentclass[pra,aps,twocolumn]{revtex4}

\usepackage{graphicx}
\usepackage{latexsym}
\usepackage{amsmath}
\usepackage{bm}

\usepackage{bm,color}

\newcommand{\be}{\begin{eqnarray}}
\newcommand{\ee}{\end{eqnarray}}

\renewcommand{\theequation}{\arabic{section}.\arabic{equation}}

\begin{document}

\title{Bosonic Analogs of Fractional Quantum Hall State in the Vicinity of Mott States 
}

\date{\today}

\author{Yoshihito Kuno} 
\author{Keita Shimizu} 
\author{Ikuo Ichinose}
\affiliation{%
Department of Applied Physics, Nagoya Institute of Technology,
Nagoya, 466-8555 Japan}

\begin{abstract}
In the present paper, the Bose-Hubbard model (BHM) with the nearest-neighbor (NN)
repulsions is studied from the view point of possible bosonic analogs of the
fractional quantum Hall (FQH) state in the vicinity of the Mott insulator (MI).
First, by means of the Gutzwiller approximation, we obtain the phase diagram of
the BHM in a magnetic field.
Then, we introduce an effective Hamiltonian describing excess particles on a MI 
and calculate the vortex density, momentum distribution and the energy gap.
These calculations indicate that the vortex solid forms for small NN repulsions,
but a homogeneous featureless `Bose-metal' takes the place of it as the NN 
repulsion increases.
We consider particular filling factors at which the bosonic FQH state is expected to form.
Chern-Simons (CS) gauge theory to the excess particle is introduced, and a 
modified Gutzwiller wave function, which describes bosons with attached flux quanta,
is introduced.
The energy of the excess particles in the bosonic FQH state is calculated using
that wave function, and it is compared with the energy of the vortex solid and Bose-metal.
We found that the energy of the bosonic FQH state is lower than that of the Bose-metal
and comparable with the vortex solid.
Finally, we clarify the condition that the composite fermion appears by using CS theory
on the lattice that we previously proposed for studying the electron FQH effect. 
\end{abstract}


\maketitle
\section{Introduction}

At present, cold atomic physics in an optical lattice is one of the most
intensively studied research field \cite{optical}.
This field opened the door for quantum simulation of
various important condensed-matter systems, 
which have been studied for a long time.
In particular by the simulation using atomic gases on an optical lattice, 
we can obtain new knowledge and view point concerning to 
the strongly-correlated many-body systems for which the conventional
methods cannot clarify the phase diagram, etc \cite{optical2}.
Recently, synthetic gauge fields mimicking uniform magnetic fields
have been created in optical lattice systems by using laser-assisted tunneling 
in a tilted optical potential \cite{uniformmg1,uniforming2}. 
The theoretical proposal for this setup was given by Jaksch and Zoller \cite{Zoller1}. 
The experiments can produce much stronger magnetic fields than those obtained
by rotating optical lattice systems \cite{rotatingmg1,rotatingmg2}. 
As a result, it is expected that a system similar to the two-dimensional (2D)
electron systems in a strong magnetic field can be produced in the atomic
gas system.

In this paper, we focus on Bose-gas systems on a 2D lattice that are
analogs of the 2D electron systems subject to a strong magnetic 
field \cite{Ezawa2,Zhang}. 
That is, we study the Bose-Hubbard model (BHM) in a strong synthetic gauge field.
In particular, we investigate the possibility of the existence of FQH state analogs in this model. 
In parallel to the experimental progress, 
there appeared many theoretical studies on the existence of integer quantum hall and FQH states 
in boson systems on the lattice \cite{Palmer,Sorensen,Hafezi,Cooper2,Hormozi,Oktel2}. 
In Refs.\cite{Sorensen,Hafezi,Mueller}, the appearance of 
the FQH-like states was suggested by calculating the overlap  
of the Laughlin wave function describing the FQH state and 
a ground-state wave function obtained by the exact diagonalization.  
Also, applying the composite fermion (CF) theory for the hard-core boson system,  
some analogous incompressible states to the FQH state on a lattice was 
studied \cite{Cooper2}.

In Ref.\cite{Oktel2}, Umucallar and Oktel gave an interesting observation that 
in the regime close to a Mott insulator (MI),
excess particles on the Mott state form a FQH-analog state at particular filling
factors of the excess particle, i.e.
the coexistence phase of the Mott state and the hard-core bosonic FQH state may exist.
Motivated by this idea, numerical studies \cite{Mueller,Natu} exhibited
the possibility of the existence of FQH-analog states in the vicinity of  the MI. 
However until new, there has been no unified view of the true ground-state 
in the vicinity of the MI in the BHM subject to a strong synthetic magnetic field.
Also, the effect of interactions has not been completely understood yet, e.g.,
how the long-range interactions, like the dipole-dipole interactions, 
change the ground-state properties.
In the optical lattice system, these interactions, as well as the on-site interactions,
are highly controllable by selecting a kind of dipolar atoms \cite{Lahaye}.

In this paper, we shall study the BHM in a strong magnetic field with and without
the nearest-neighbor (NN) interactions.
In particular, we investigate properties of the ground-states in the vicinity of 
the MI and effects of the NN interaction on them by using both the Chern-Simons (CS) 
theory \cite{Ezawa2,Zhang,Tong} and a numerical Gutzwiller method \cite{TDGW1,TDGW2}.  
For commensurate magnetic fields and particle fillings, the  Gutzwiller 
method show that a vortex solid
form for weak NN repulsions, whereas for relatively strong NN repulsions,
a featureless homogeneous state (we call Bose-metal) takes the place of the vortex solid. 
To investigate the possibility of the bosonic FQH state, we develop the method that
we call CS-Gutzwiller wave function.
In that wave function, an integer number of flux quanta are attached to 
each (excess) particle as described by the CS theory.
The energy of the states described by the CS-Gutzwiller wave function is
compared with that of the vortex solid and Bose-metal, and we obtain interesting results.

This paper is organized as follows. 
In Sec.\ref{CII}, we outline the target BHM and introduce an effective Hamiltonian 
that describes the  excess particle in the vicinity of Mott states.
In Sec.\ref{CIII}, we carry out the Gutzwiller numerical method for the BHM in a
synthetic gauge filed, and observe the ground-states, vortex configurations
, momentum distributions and an excitation gap on the ground-state.
Next, we apply the lattice CS theory to the BHM and 
analyze an excitation spectrum and gap by using the Bogoliubov theory in Sec.\ref{CIV}.
In Sec.\ref{CV}, we construct the CS-Gutzwiller numerical method and apply it to 
the excess particle Hamiltonian. 
Then, we estimate the energy of the CS-Gutzwiller ground-states and compare it
to the energy of the vortex solid and Bose-metal obtained in Sec.\ref{CIII}. 
In Sec.\ref{CVI}, from the view of the CF theory, we discuss 
the excess particle system and show the condition that the CF picture appears
as low-energy excitations.
There, the gauge-theoretical consideration plays an importance role.
Finally in Sec.\ref{CVII}, we propose an experimental method to detect 
a ground-state excitation gap for the FQH-analog state, and conclude
the present study.
In the appendix, we consider the practical cold atomic systems and estimate the on-site
and NN repulsions.
There, the NN repulsion between atoms is provided by the dipole-dipole interaction of atoms.

\section{Model and effective Hamiltonian in the vicinity of Mott plateaus} \label{CII}

In this paper, we consider 2D bosonic gases described by the 
BHM in a strong magnetic field with NN repulsions. 
The Hamiltonian $H_{\rm BHM}$ of the BHM on the 2D square lattice is given as 
\begin{eqnarray}
H_{{\rm BHM}}&=&-J\sum_{\langle i,j\rangle} 
(a^{\dagger}_{i} a_{j}e^{iA_{ij}}+\mbox{H.c.})  
+\sum_{i}\frac{U}{2} n_i(n_{i}-1) \nonumber \\
&&+V\sum_{\langle i,j\rangle}n_in_j-\mu \sum_{i}n_{i},
\label{BH}
\end{eqnarray}
where, $a_i \ (a^\dagger_i)$ is the boson annihilation (creation) operator at site $i$ and 
$n_{i}= a^{\dagger}_{i} a_{i}$. 
$\langle i,j \rangle$ denotes a pair of NN sites.
The parameter $J$ is the NN hopping amplitude, 
$U$ and  $V$ are the on-site and NN repulsions, respectively. 
In real experiments, a ratio $V/U$ is highly controllable
and can be a fairly large value to a certain extent, see the discussion in appendix A. 
In this paper, we take the value of $V/U$ up to $\sim 0.3$.
The vector potential $A_{ij}$ represent a uniform magnetic field and satisfies
$\sum_{\rm plaquette} A_{ij}=2\pi f$ with a parameter $0\leq f \leq 1$.
In this paper, we mostly focus on the case $f={1 \over 2}$ and sometimes 
$f={1 \over 3}$, although a generalization
to the case $f=t/s$ ($s$ and $t$ co-prime integers) is rather straightforward.

\begin{figure}[t]
\centering
\includegraphics[width=8cm]{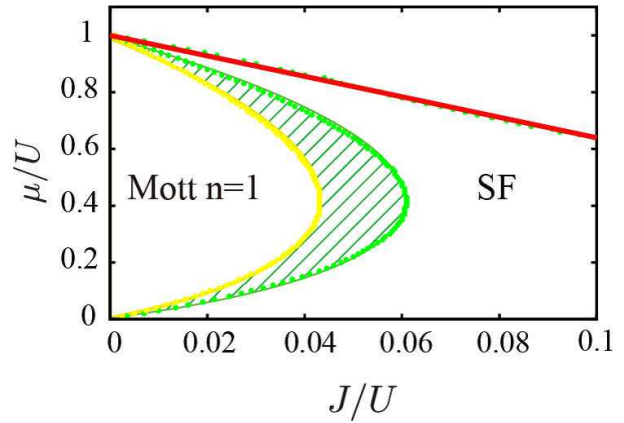}
\caption{(Color online) Obtained phase diagram of the BHM subject to 
a strong magnetic field.
The yellow line is the phase boundary separating the MI and SF states
for vanishing magnetic field $f=0$ obtained by the Gutzwiller numerical method. 
The green line is an elongated phase boundary as a result of 
the applied magnetic flux per plaquette, $2\pi f=\pi$. 
The red line represents the states with the average particle density ${\rho}=  1.25$.
}
\label{Fig1}
\end{figure}

As is well known,
the BHM has the MI and superfluid (SF) phases, whose phase boundary forms lobes.
Figure \ref{Fig1} shows the phase diagram obtained by our Gutzwiller numerical method
that we shall explain in later section. 
In general the MI phase is enhanced by the magnetic field, i.e.,
the lobes elongate compared to the case without the magnetic field.
The phase diagram in Fig.\ref{Fig1} is in good agreement with the previous 
works in Refs.\cite{Oktel1,Oktel2,Mueller}.
In what follows, we shall study the BHM in the vicinity 
of the MI of the integer particle filling.

To this end, we consider an effective Hamiltonian that describes particles
in the vicinity of the Mott lobes.
In Ref.\cite{Oktel1}, it was discussed that in the regime of a uniform filling 
$\rho=n+\rho_{\rm ep} \ (n=\mbox{a positive integer}, \ 0<\rho_{\rm ep}\ll 1)$
the total boson system is divided into two part: 
MI part of the filling factor $n$ and the excess particle part of particle
density $\rho_{\rm ep}$.
In this picture, the excess particles are moving on the solid-like MI. 
However
as the MI particle and the excess particle are the same kind of particles, the quantum
symmetrization of the  Bose particle has to be imposed on the many-particle 
quantum state.
  
In the excess particle sector, the hopping parameter of the excess particle changes as  
$J\rightarrow J(n+1)$, and the on-site interaction $U/2\rightarrow U$
from the original ones \cite{Oosten,Oktel2}, whereas $V$ is intact.
Then, the effective Hamiltonian of the excess particle $H_{\rm eBHM}$is given as,
\begin{eqnarray}
H_{{\rm eBHM}}&=&-J(n+1)\sum_{\langle i,j\rangle} 
(c^{\dagger}_{i} c_{j}e^{iA_{ij}}+\mbox{H.c.})  \nonumber \\
&&+\sum_{i}U n_{ci}(n_{ci}-1)
+V\sum_{\langle i,j\rangle}n_{ci}n_{cj}\nonumber\\
&& -\tilde{\mu} \sum_{i}n_{ci},
\label{eBH}
\end{eqnarray}
where $c_{i}(c^{\dagger}_{i})$ is an annihilation (creation) operator of excess boson
on site $i$, the number operator $n_{ci}= c^{\dagger}_{i} c_{i}$, and parameter 
${\tilde \mu}$ is the chemical potential for the excess particle. 

The above results are derived by the following consideration.
The MI state with the filling $n$ is given as,
\begin{eqnarray}
|{\rm MI}\rangle = \prod_{i=1}^N(a^{\dagger}_{i})^n|0\rangle,
\end{eqnarray}
where $N$ is the total number of lattice site, and $|0\rangle$ is the vacuum sate, 
which includes no particle.
The state $|{\rm MI}\rangle$ is a base state on considering the excess particle
Hamiltonian, i.e., its plays a role of the vacuum state of $H_{\rm eBHM}$.
Next, we consider one-particle creation on the state $|{\rm MI}\rangle$ 
\begin{eqnarray}
|1{\rm particle}\rangle \equiv \biggl(\sum_{i}\Psi^{(k)}_i a^{\dagger}_{i} \biggr)
|{\rm MI}\rangle,
\end{eqnarray}
where $\Psi^{(k)}_i$ is a wave function of the particle. 
Then, let us consider the hopping energy of this state $|1{\rm particle}\rangle$.
To simplify the discussion, we consider a single-particle wave function 
in the Hofstadter butterfly \cite{Hofstadter} for $\Psi^{(k)}_i$ with energy $\epsilon_k(f)$.
Please notice that $\{\Psi^{(k)}\}$ form a complete set of the Hilbert space
of the state vectors. 
Then applying the original hopping term to the state $|1{\rm particle}\rangle$,
we have
\begin{eqnarray}
&&\biggl( \sum_{l,j}t_{lj}a^{\dagger}_l a_{j}\biggr)|1 {\rm particle}\rangle \nonumber\\
\rightarrow&&  J\epsilon_{k}(f)(n+1) \biggl(\sum_{i}\Psi^{(k)}_i a^{\dagger}_{i} \biggr)
|\rm MI\rangle \nonumber\\
&&= J\epsilon_{k}(f)(n+1)|1 {\rm particle}\rangle,
\label{1particle1}
\end{eqnarray} 
where $t_{lj}$ stands for general hopping amplitudes and $t_{lj}=Je^{iA_{lj}}$
in the present case.
From Eq.(\ref{1particle1}), we can see that the hopping energy of the excess
particle is given by $(n+1)t_{ij}$.  
This result is in agreement with the previous results by analytical calculations of the excitation spectrum 
\cite{Oosten,Sengupta} and the numerical study \cite{Oktel1}.
In intuitive picture, the above result can be understood as follows.
There are $(n+1)$ bosons and they are all equal footing
and any of them can hop to a NN site, then the hopping amplitude of the excess
particle is $(n+1)$-fold of the original one.

Next, we consider the on-site interaction energy for the excess particle. 
To begin with, we consider one-particle on-site energy deviation from the MI state.
To this end, we put $\Psi^{(k)}_i=\delta_{ij}$, which is another complete set of the
wave functions.
It is rather straightforward to calculate
\begin{eqnarray}
&&\langle 1{\rm particle}|\frac{U}{2}{\hat n}^{2}|1{\rm particle}\rangle - \langle {\rm MI}|\frac{U}{2}{\hat n}^{2}|{\rm MI}\rangle \nonumber\\
&& =\frac{U}{2}(2n+1).
\end{eqnarray} 
Thus, the on-site energy of the state $|1 {\rm particle}\rangle$ is $\frac{U}{2}(2n+1)$.
Similarly, we can consider two-particle on-site energy deviation from the MI state.
\begin{eqnarray}
&& |2 {\rm particle}\rangle\equiv (a^{\dagger}_{j} )
|1 {\rm particle}\rangle, \nonumber\\
&&\langle 2 {\rm particle}|\frac{U}{2}{\hat n}^{2}|2{\rm particle}\rangle - \langle {\rm MI}|\frac{U}{2}{\hat n}^{2}|{\rm MI}\rangle \nonumber\\
&&\hspace{1.7cm} = U(2n+2).
\end{eqnarray} 
From the above results, the two-body interaction energy of the excess particle is
obtained as, 
$$U(2n+2)-2\frac{U}{2}(2n+1)=U.$$
Similar discussion on the NN repulsion $V\sum_{\langle i, j\rangle}n_in_j$
shows that the NN repulsion of the excess particle remains the same.
In this way, the effective Hamiltonian $H_{\rm eBHM}$ in Eq.(\ref{eBH})
is derived.

In the rest of this paper, we shall study the model $H_{\rm eBHM}$ in Eq.(\ref{eBH})
by means of the numerical as well as analytical methods.


\section{Numerical Study by Gutzwiller approximation} \label{CIII}
\setcounter{equation}{0}

In this section, we introduce the Gutzwiller approximation \cite{TDGW1,TDGW2} 
that is useful for studying equilibrium states in the strong interaction regime like 
the MI and its vicinity. 
Then in this section by means of the Gutzwiller approximation, 
we study the system of the total particles described by the Hamiltonian 
$H_{\rm BHM}$ in Eq.(\ref{BH}).
In the practical calculation, we mostly focus on the case $f={1 \over 2}$ and
$n=1$, and the density of excess particle per site $\rho_{\rm ep}={1 \over 4}$, 
i.e., the filling fraction of excess particle 
$\nu_{\rm ep}={\rho_{\rm ep} \over f}={1 \over 2}$,
i.e. total mean density $\rho=1.25$.

\subsection{Gutzwiller Method}

We first introduce a Gutzwiller-wave function constructed from
the particle number bases of each site $i$, 
\begin{equation}
|\Psi_{\rm GW}\rangle = \prod_{i=1}^N
\biggl( \sum^{n_{c}}_{n=0} f^{i}_{n}|n\rangle_{i}\biggr), 
\label{Gwf}
\end{equation}
where $N$ is the number of the lattice sites, $n_{c}$ is a maximum particle number 
at each site that is a parameter in the Gutzwiller approximation,
and the coefficients $\{f^{i}_{n}\}$ are variational parameters, which are to be determined
by solving the decoupled Hamiltonian given below.
As the variational parameters $\{f^{i}_{n}\}$ are defined on each site,
the total number of parameter is $N^{n_{c}}$.

In order to obtain $\{f^{i}_{n}\}$ for the ground-state wave function, we employ 
a mean-field type approximation, i.e., we decouple the hopping and the NN repulsion
terms in $H_{\rm BHM}$ in Eq.(\ref{BH}) and derive a single-site Hamiltonian
$h_{{\rm BHM} i}$. 
Then, we introduce an order parameter of the SF, i.e., Bose-Einstein condensation (BEC),
\begin{equation}
\Psi_{i}\equiv \sum^{n_{c}}_{n_d=1}\sqrt{n_d}f^{* i}_{n_d-1}f^{i}_{n_d},
\label{opBEC}
\end{equation}
From Eq.(\ref{Gwf}), it is obvious that 
$\langle \Psi_{\rm GW}|a_i |\Psi_{\rm GW}\rangle=\Psi_i$.
With $\Psi_i$, the local Hamiltonian is given as,
\begin{eqnarray}
h_{{\rm BHM} i}&=&-J\sum_{j\in {\rm iNN}}\biggl(a^{\dagger}_{i}
e^{iA_{ij}}\Psi_{j}+\mbox{h.c.}\biggr)
\nonumber\\
 &&+\frac{U}{2}n_{i}(n_{i}-1)+Vn_i\Big(\sum_{j\in i{\rm NN}}n_j\Big) \nonumber \\
&&-\mu n_{i},
\end{eqnarray}
where $j\in i{\rm NN}$ denotes the NN sites of site $i$.
From the local Hamiltonian $h_{{\rm BHM} i}$,
the site-energy $E_i$ is estimated as follows by using the wave function 
$|\Psi_{\rm GW}\rangle$,
\begin{eqnarray}
E_{i}&=&\langle\Psi_{\rm GW}|h_{{\rm BHM} i}|\Psi_{\rm GW}\rangle\nonumber\\
&=&\sum^{n_{c}}_{n_d=0}\biggl[-J\sum_{j}\biggl(\sqrt{n_{d}}f^{i}_{n_{d}-1}
e^{iA_{ij}}\Psi_{j}
\nonumber\\
&+&\sqrt{n_d+1}f^{i}_{n_d+1}e^{-iA_{ij}}\Psi^{*}_{j}\biggr){f^{* i}_{n_d}} \nonumber\\
&+&\biggl(\frac{U}{2}n_{d}(n_{d}-1)-\mu n_{d}\biggr) f^{i}_{n_{d}} {f^{* i}_{n_d}} 
\nonumber \\
&+&Vn_df^{i}_{n_{d}} {f^{* i}_{n_d}} \Big(\sum_{j\in i{\rm NN}}\langle n_j \rangle\Big)
\biggr],
\label{local_energy}
\end{eqnarray}
where $\langle n_j \rangle$ is the expectation value of $n_j$.
This local mean-field energy $E_{i}$ and the mean field $\Psi_{i}$ form a 
self-consistent equation. 
By using an iterative process \cite{Oktel2,Rigol}, the total energy 
$E=\sum_iE_i$ can be minimized and both the corresponding variational 
parameters ${f^i_{n_d}}$ 
and order parameter $\Psi_{i}$ are obtained simultaneously.
In our practical calculation, we mostly fix the truncated particle number $n_{c}=7$ as
this value is expected to be large enough to capture physics in our target regime. 
We have verified this expectation by varying the value of $n_c$ for 
some specific quantities.
Also, most of the calculations were performed for the linear system size $L=12$
with the periodic boundary condition. 

\begin{figure*}[t]
\centering
\includegraphics[width=15cm]{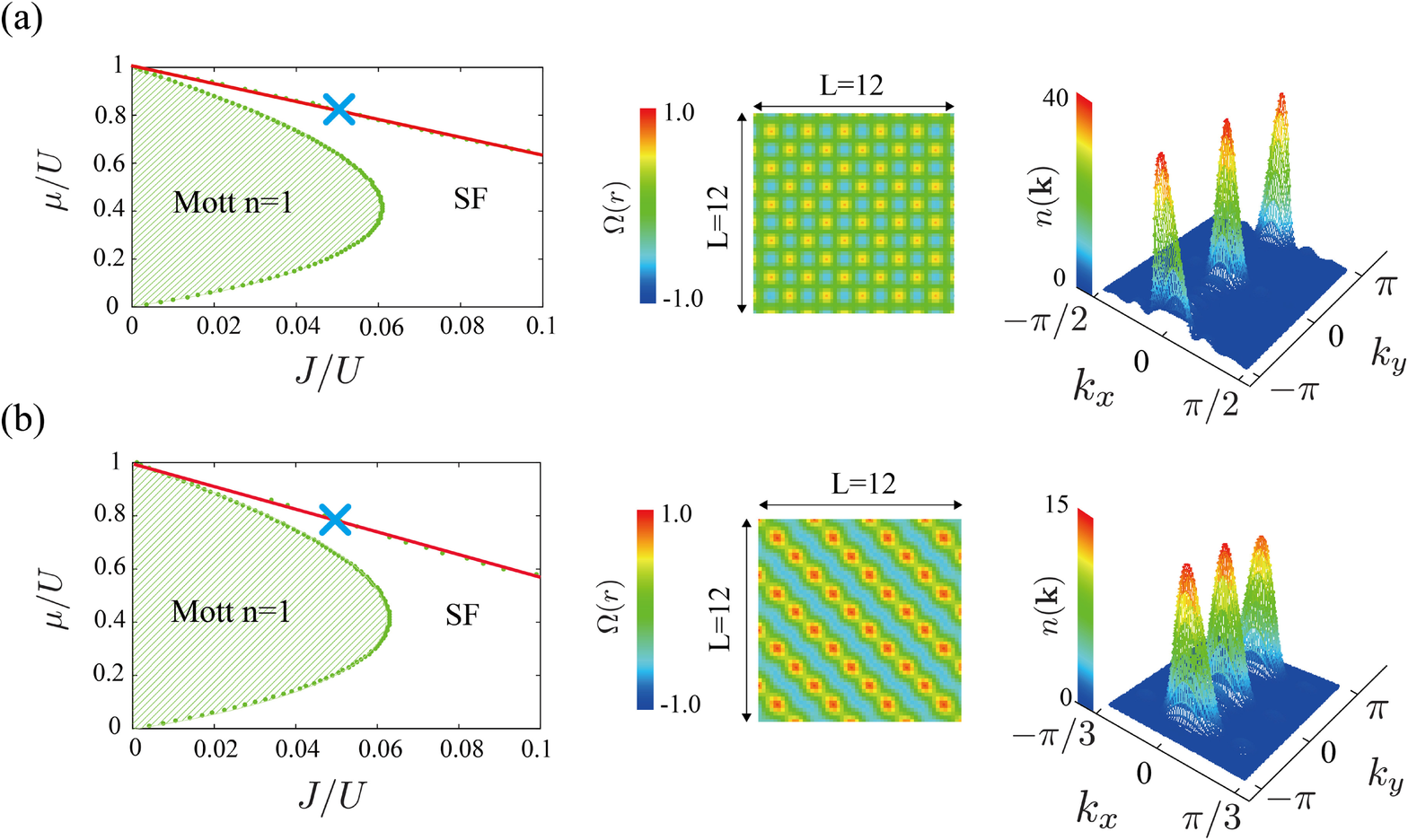}
\caption{(Color online) Numerical results for the $V/U=0$ and $J/U=0.05$ case. 
The upper panels (a) correspond to the $f=1/2$ case. 
The phase diagram in the left most panel exhibits the point of $\nu_{\rm ep}=1/2$ 
by the cross ${\bf \times}$
at which the measurements of vortex density $\Omega(r)$, and the momentum 
distribution $n({\bf k})$ were performed.
The middle and right panels show the calculations of $\Omega(r)$ and $n({\bf k})$,
respectively.
The lower panels (b) correspond to the $f=1/3$ case ($\nu_{\rm ep}=1/2$).
In both cases, the results obviously show that the stable vortex solid forms.
}
\label{V=0_0.2_vortex}
\end{figure*} 

\begin{figure*}[t]
\centering
\includegraphics[width=15cm]{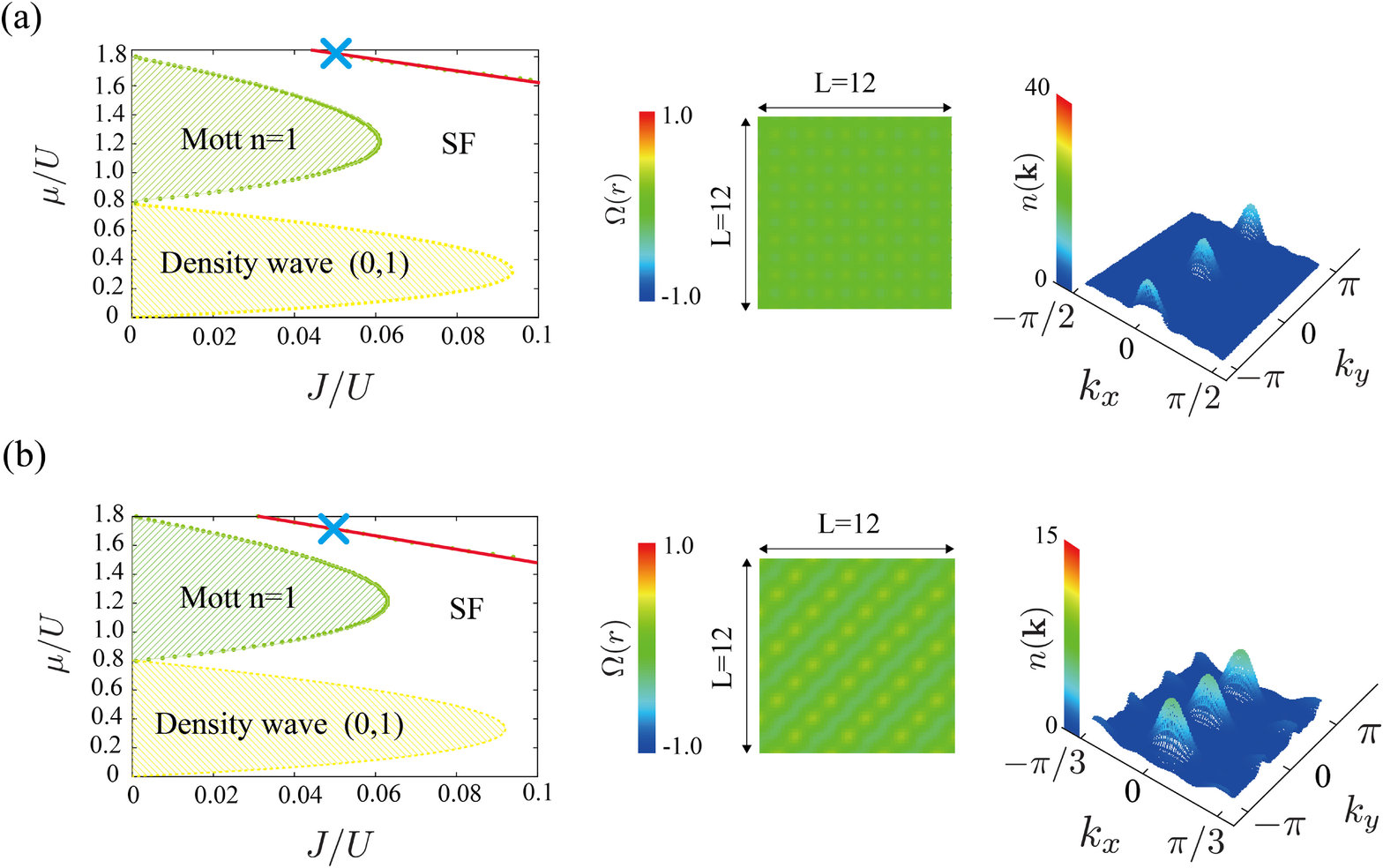}
\caption{(Color online) Numerical results for the $V/U=0.2$ and $J/U=0.05$ case. 
As in Fig.\ref{V=0_0.2_vortex}, the upper panels correspond to the $f=1/2$ case 
and the lower panels to the $f=1/3$ case.
Calculations of  $\Omega(r)$ and $n({\bf k})$ are shown.
In both cases, the signals of the vortex solid are weakened, in particular, 
in the $f=1/2$ case. 
}
\label{V=0.2_vortex}
\end{figure*} 

\subsection{Numerical results in the vicinity of the Mott plateaus}

In solving the Gutzwiller-wave equations practically,
there is a point that has to be taken into account carefully.
Solution to the Gutzwiller-wave equation usually depends on an initial condition 
\cite{TWA0}.
That is, solution sometimes goes to local minimum and does not reach 
the true ground-state due to a large number of the variational parameters 
$\{f^{i}_{n_d}\}$.
To overcome this difficulty, we performed the calculations by varying the 
initial configurations in various ways, and searched solutions of the lowest-energy state 
by trial and error.

\begin{figure}[h]
\centering
\includegraphics[width=7cm]{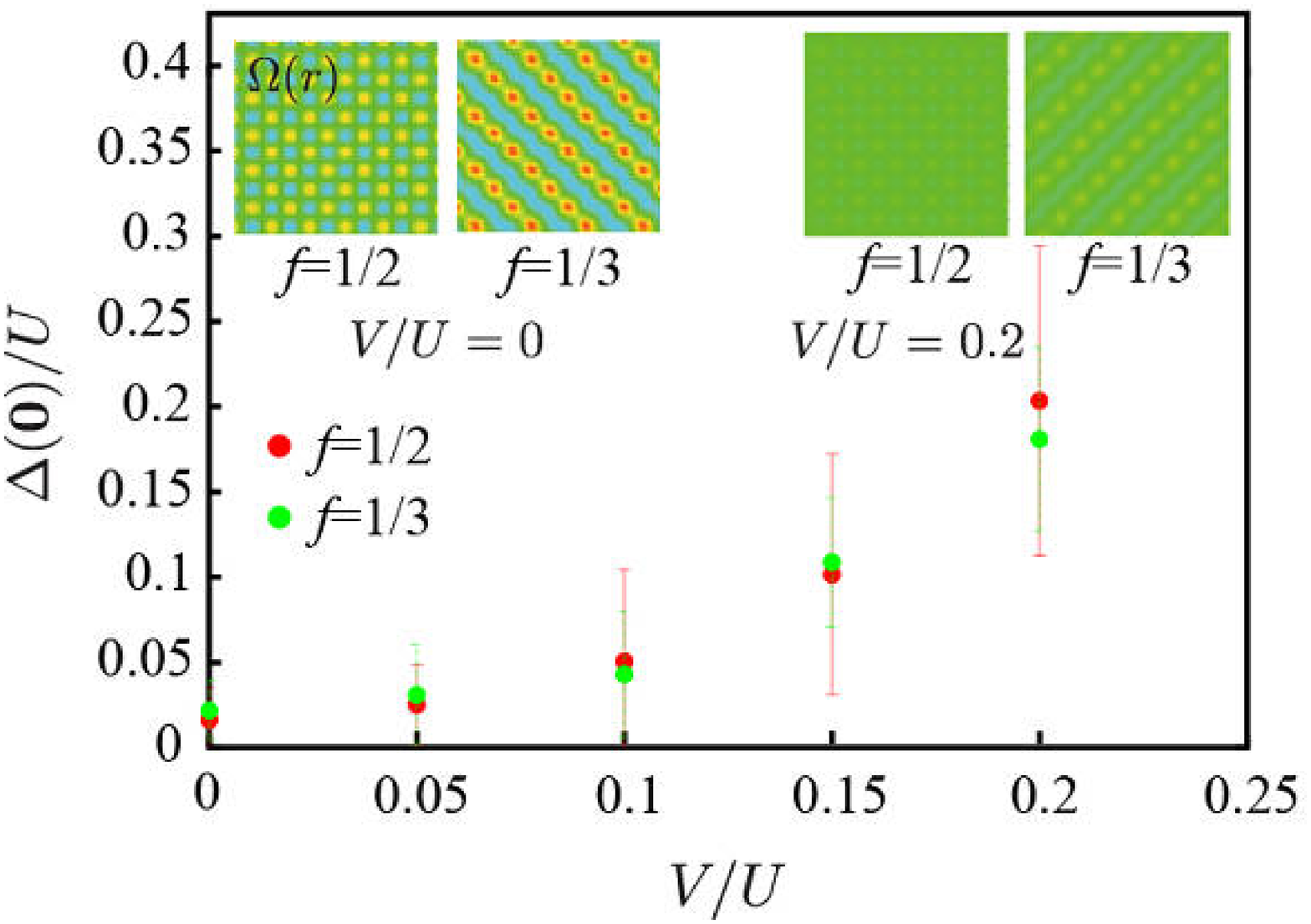}
\caption{(Color online) Energy gaps calculated by the single-mode approximation
for the states described by the Gutzwiller-wave functions in the case $J/U=0.05$.
Applied magnetic field is $f=1/2$ and $1/3$ and $\rho\simeq 1.25$.
As the NN repulsion $V$ increases, the energy gap increases from the
vanishing value.
At $V=0$, the stable vortex solid forms as the density profile in the inset indicates.
As a result, the gapless Nambu-Goldstone boson exists.
On the other hand for $V/U=0.2$, the vortex solid melts and the SF is destroyed,
and then, the excitations acquire a gap.
We took 50 samples in each measurement because 
the value of the excitation gap depends on initial values of $\{\Psi_{i}\}$.
}
\label{gap_V}
\end{figure}

To study the ground-state physical properties, we calculated vortex configurations, 
the density momentum distribution and also the energy gaps. 
Vorticity $\Omega(r)$ at the dual lattice site $r$ is given as 
\begin{eqnarray}
&&\Omega(r) = \frac{1}{2\pi}\sum_{\mu,\nu}\epsilon_{\mu \nu}\nabla_{\mu}J_{i, \nu},  
\nonumber   \\
&&\epsilon_{12}=-\epsilon_{21}=1, \; \epsilon_{11}=\epsilon_{22}=0, \nonumber \\
&&\nabla_\mu J_{i,\nu}=J_{i+\mu,\nu}-J_{i,\nu},
\label{Omega}
\end{eqnarray}
where $J_{i,\nu}$ is a current of $\Psi_i$ in the $\nu$-direction
defined by the hopping term in the BHM, and explicitly given as 
$J_{i,\nu}=\frac{1}{4}\sin(\theta_{i+\nu}-\theta_{i})$ for $\Psi_i=\sqrt{\rho_{i}}e^{i\theta_{i}}$.
The quantity $\Omega(r)$ measures a density of pinned quantized vortices in the SF
that may arise by the applied magnetic field. 
The momentum distribution is given as
$$
n({\bf k})=\frac{1}{N}\sum_{i,j}\langle a^{\dagger}_{i}a_{j}\rangle e^{i{\bf k}
\cdot ({\bf R}_{i}-{\bf R}_{j})}
=\frac{1}{N}\sum_{i,j}\Psi^{*}_{i} \Psi_{j}e^{i{\bf k}\cdot ({\bf R}_{i}-{\bf R}_{j})},
$$
in the Gutzwiller approximation.
Quantity $n({\bf k})$ clarifies the momentum ${\bf k}$ at which the BEC takes place. 
As the analytical study in Ref.\cite{Sinha} shows, a condensate with a non-vanishing 
${\bf k}$ is expected to form. 
In the following numerical study, we shall verify the existence of such a condensation
for certain parameter regions.

First, we consider the case of the vanishing NN repulsion $V=0$, and
show the numerical results.
As we stated above, the magnetic flux per plaquette is $2\pi f=\pi$ and the density
of the (excess) particle $\rho=1.25 \ (\rho_{\rm ep}=0.25)$, i.e., the filling factor
of the excess particle $\nu_{\rm ep}={0.25 \over 0.5}={1\over 2}$.

The upper-left panel in Fig.\ref{V=0_0.2_vortex} shows the phase diagram in the 
$(J/U-\mu/U)$-plane and also the line $\rho=1.25$ is indicated.
The cross symbol on the line $\rho=1.25$ exhibits the parameter
($J/U=0.05, \ \mu/U=0.8$) on which 
we calculated the vortex density $\Omega(r)$ and the density momentum distribution 
$n({\bf k})$.
By the application of the magnetic field, the MI phase is elongated.
The upper-middle panel shows 
that vortices are generated and they crystallize and form a solid pattern
as a result of the pinning by the lattice and inter-vortex repulsion. 
In this vortex solid state, the momentum distribution $n({\bf k})$ clearly exhibits 
a BEC at a finite momentum in the first magnetic Brillouin zone. 
The appearance of 
the same vortex solid pattern was shown for {\it deep SF states}
by the previous work using large scale Monte Carlo simulations
\cite{Kuno_Nakafuji_Ichinose}.
The numerical results $\Omega(r)$ and $n({\bf k})$ indicate that the BEC forms
even though the condensation appears at non-vanishing ${\bf k}$ points. 
The state breaks the global U(1) symmetry of the phase rotation and
as a result, the ground-state has a gapless excitation as we show later on. 
This state is the genuine SF.   

Similar results were obtained for the case $\nu_{\rm ep}={1 \over 2}$ with
$f={1 \over 3}$ and $\rho=1+ {1\over 6}$.
See the lower panels in Fig.\ref{V=0_0.2_vortex}.
The pattern of the vortex solid is the same with that observed in the previous
work for the deep SF state with $f={1 \over 3}$ \cite{Kuno_Nakafuji_Ichinose}.

Let us study the effects of the NN repulsion.
We studied the case $V/U=0.2$, and the obtained phase diagrams are shown in 
Fig.\ref{V=0.2_vortex}.
Finite NN repulsion shifts the MI-SF boundary and also the density-wave (DW) state
with the density $\rho={1 \over 2}$ appears, in which the particle density at the
even (odd) sublattice is unity (vanishing) or vice versa.
We denote this state as $(0,1)$ DW in Fig.\ref{V=0.2_vortex}.
In order to study the case of the filling fraction $\nu_{\rm ep}={1 \over 2}$ near 
the MI with $\rho=1$, we adjusted the chemical potential properly.
The values of $\mu/U$ and $J/U$ for the numerical study are indicated in 
the phase diagram in Fig.\ref{V=0.2_vortex} by the cross symbol.

The calculated local vortex density $\Omega(r)$ and the momentum distribution 
$n({\bf k})$ are shown in Fig.\ref{V=0.2_vortex}.
Interestingly enough, the calculation of $\Omega(r)$ shows that the vortex 
solid melts, and a featureless state takes the place of the vortex solid.
This result is confirmed by the calculation of $n({\bf k})$.
The result in Fig.\ref{V=0.2_vortex} exhibits the smearing of peaks 
that existed in the case of $V=0$.
We also verified that there exist no phase coherence of $\Psi_i$, i.e., 
$\Psi_i$ substantially changes spatially and also under the local update 
of $\{f^i_n\}$.
From the above observation, we conclude that the obtained quantum state
for $V/U=0.2$ is {\em not} the SF, and the U(1) symmetry of the phase rotation
is preserved.
It is also obvious that the state under consideration is not the MI, and therefore
it may be called `Bose-metal'.

In order to verify the above conclusion, we calculated the energy gap from
the obtained ground-states.
In the single-mode approximation \cite{Feynman}, 
the excitation spectrum in ${\bf k}$-space is given by
\begin{eqnarray}
\Delta({\bf k})=
\frac{\langle \Psi_{\rm GS}|\rho^{\dagger}_{{\bf k}}(\hat{H}-\epsilon_{0})
\rho_{{\bf k}}|\Psi_{\rm GS}\rangle}{\langle \Psi_{\rm GS}|\Psi_{\rm GS}\rangle},
\label{exc}
\end{eqnarray}
where 
$|\Psi_{\rm GS}\rangle$ denotes the ground-state wave function
and $\epsilon_{0}$ is its energy.
The density operator $\rho_{{\bf k}}$ is given as follows in the second-quantized
representation, 
$$
\rho_{{\bf k}}=\sum_{\ell}e^{i{\bf k}\cdot {\bf R}_{\ell}}\hat{n}_{\ell}, \; \;
\hat{n}_{\ell}=a^\dagger_\ell a_\ell.
$$ 
By taking ${\bf k}\rightarrow {\bf 0}$, $\Delta({\bf 0})$ gives an excitation gap from the ground-state. 
We apply the above formulation to the Gutzwiller-wave function, i.e., 
the ground-state wave function $|\Psi_{\rm GS}\rangle$ is taken to
the Gutzwiller ground-state wave function $|\Psi_{\rm GW}\rangle$
obtained for the parameters from $V/U=0$ to $V/U=0.2$.

Figure \ref{gap_V} shows the excitation gaps of the ground-states. 
The gap is vanishingly small for $0<V<0.1$ whereas it starts to increase as $V$
increases from $0.1$.
This result indicates that the BEC realizes for small $V$ but for $V>0.1$,
another gapped state appears as the above consideration suggests.
Recently, similar gapped states have been reported in Ref.\cite{Natu}, 
where a cluster type numerical mean-field method was used for the numerical
calculation.    

It is interesting to search another state that cannot be described by the
site-factorized Gutzwiller-wave function in Eq.(\ref{Gwf}).
A candidate of such states is the bosonic FQH state. 
In Sec.V, we shall show that such a state can be described
by the modified Gutzwiller-wave function based on the idea of the 
flux attachment to particle, and in fact, it can be a candidate of the 
ground-state with strong correlations.
Before going into the details of the calculation, we review the lattice 
CS gauge theory for the bosons on the lattice in the following section.

\section{Lattice Chern-Simons theory for excess particle and 
excitation gap of composite boson}  \label{CIV}
\setcounter{equation}{0}

\begin{figure}[t]
\centering
\includegraphics[width=6cm]{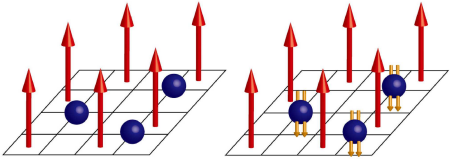}
\caption{(Color online) Chern-Simons theory of lattice bosons in a strong
magnetic field.
Each boson is attached an even number of the flux quanta (orange arrow) of the CS gauge field.
The external magnetic field (red arrow) is canceled out on the average by the CS gauge field. 
}
\label{CStheory}
\end{figure}

We apply the CS theory to the excess particle Hamiltonian (\ref{eBH}).
The CS theory succeeded in describing
the FQH state in 2D electron system \cite{Ezawa2,Zhang,Tong}.
One of the authors previously introduced and formulated the lattice version of 
the CS theory for 2D lattice fermion system \cite{IchinoseMatsui}, and
this formulation is well suited for study of the present boson system.
See Fig.\ref{CStheory}.

By using this formalism, we transform the original excess boson operator $c_i$ in 
Eq.(\ref{eBH}) to another particle operator $b_i$, which we call CS particle, 
by attaching $(\nu_{\rm ep})^{-1}$-magnetic flux quanta to $c_i$,
\begin{eqnarray}
&&c_{i}=U_{i}b_i, \nonumber \\
&&U_i=\exp\Big[ i\nu_{\rm ep}^{-1}\sum_{r'}\theta(i,r')(c^\dagger_{i'}c_{i'})\Big],
\label{CSboson}
\end{eqnarray}
where $r \ (r')$ denotes a site of the dual lattice paired to site 
$i \ (i')$ of the original 
lattice as before, and $\theta(i,r')$ is the azimuthal angle function on the lattice.
As we consider the case in which $\nu^{-1}_{\rm ep}=$ an integer, 
the transformation 
(\ref{CSboson}) is well-defined.
Please notice $c^\dagger_i c_i=b^\dagger_i b_i$, and therefore
\begin{eqnarray}
H_{\rm eBHM}&=&-J(n+1)\sum_{\langle i,j\rangle}(b^\dagger_i W^\dagger_i W_{j}b_{j}
+\mbox{h.c.}) \nonumber\\
&+&\sum_{i}U(b^{\dagger}_{i} b_{i}-1)b^{\dagger}_{i} b_{i}
+V\sum_{\langle i,j\rangle}b^{\dagger}_{i} b_{i}b^{\dagger}_{j} b_{j} \nonumber\\
&&-{\tilde \mu}\sum_{i}b^{\dagger}_{i} b_{i} 
\label{BHCS1} \\
W_i&=&\exp\Big[ i\nu_{\rm ep}^{-1}\sum_{r'}\theta(i,r')(b^\dagger_{i'}b_{i'}
-\rho_{\rm ep})\Big].\nonumber
\label{BHCS2}
\end{eqnarray}
Here we have employed the symmetric gauge for 
$A_{i,\mu}\equiv A_{ij=i+\mu}$, and used the identity
\begin{equation}
2\pi\epsilon_{\mu\nu}\nabla_\nu G(r,r')=\nabla_\mu\theta(i,r'),
\label{angleG}
\end{equation}
where $G(r,r')$ is the two-dimensional lattice Green function, i.e.,
$\sum_{\mu=1,2}\nabla^2_\mu G(r,r')=-\delta_{rr'}$.
The CS gauge theory can be constructed for the system in
Eq.(\ref{BHCS1}) in the Lagrangian formalism,
but here we only discuss the possible mean-field solution of the ground-state
and low-energy excitations of the above system.
Hereafter as a example, we consider the mean excess particle density 
$\rho_{\rm ep}={1 \over 4}$ and the magnetic field $f={1 \over 2}$, and 
as a result, $\nu_{\rm ep}={1 \over 2}$. 
The left panel in Fig.\ref{V=0_0.2_vortex} (a) indicates the line 
$\rho_{\rm ep}={1 \over 4}$. 
In the case $\nu_{\rm ep}={1 \over 2}$, two flux quanta is attached to 
one excess particle, and then the CS particle is bosonic, i,e., composite boson (CB). 
Also, we shall discuss a CF picture in Sec.\ref{CFpicture}

\begin{figure*}[t]
\centering
\includegraphics[width=16.5cm]{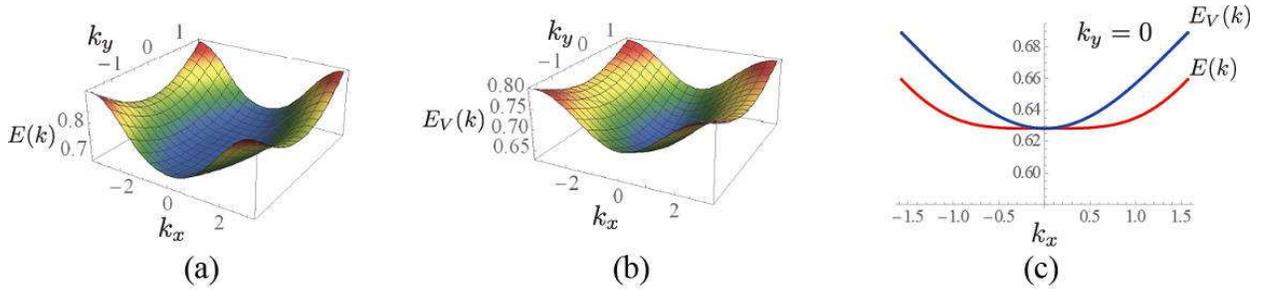}
\caption{(Color online) Bogoliubov excitation spectrum of $\tilde{J}/U=0.1$ 
(a) for $V/U=0$ and (b) $V/U=0.2$. 
(c) Dispersion relations for $V/U=0$ and $V/U=0.2$, $E(k)$ and $E_{V}(k)$,
in the plane $k_{y}=0$.
$E(0)=E_V(0)$, however, $E_V(k)>E(0)$ for ${\bf k}\neq 0$.
This indicates the stability of the CB picture for $V>0$.
}
\label{CB_excitation}
\end{figure*} 

If the CB forms a BEC in the system $H_{\rm eBHM}$ of Eq.(\ref{BHCS1}), 
a bosonic analog of the FQH state is realized.
In this case,  the expectation value of the CB operator is
the density $\rho_{\rm ep}$, which is uniform on the present lattice system
as the external magnetic field is canceled by the CS gauge field at temperature
$T=0$, i.e., $\langle W_i \rangle=1$. 
Under this assumption, we can calculate the energy gap from the Hamiltonian (\ref{BHCS1}). 
We impose hard-core boson constraint, and then we drop the on-site interaction 
term with the coefficient $U$. 
This result comes from two reason; 
excess particle density is dilute and we consider the large-$U$ regime.
We first put $V=0$ for simplicity and the effect of the NN
repulsion will be studied afterward.

We introduce a quantum fluctuation $\eta_{i}$ from the condensation of 
the CB operator.
\begin{eqnarray}
b_{i}=\sqrt{\rho_{\rm ep}}+\eta_{i},
\label{CBDC}
\end{eqnarray}
The uniform density $\rho_{\rm ep}$ determines the chemical potential
$\tilde{\mu}$.
Substituting the uniform condensed variable $b_{i}\rightarrow b$, 
the Hamiltonian $H_{{\rm eBHM}}$ reduces to
\begin{eqnarray}
H_{{\rm eBHM}}\rightarrow E_b = -4\tilde{J}|b|^2 - \tilde{\mu}{|b|^2},
\end{eqnarray}
where we have put $\tilde{J}\equiv J(n+1)$.
From this mean field energy, the chemical potential is determined as, 
\begin{eqnarray}
\frac{d}{d{b^{*}}}E_b\biggl{|}_{b={\sqrt{\rho_{\rm ep}}}}=0 
\rightarrow \;\; \tilde{\mu}=-4\tilde{J}.
\label{tilmu}
\end{eqnarray}

From Eqs.(\ref{CBDC}) and (\ref{tilmu}), we derive the effective Hamiltonian
of $\eta_i$.
First, the hopping term in Eq.(\ref{BHCS1}) is rewritten as, 
\begin{eqnarray}
\sum_{i,j}(b^{\dagger}_{i} W^{\dagger}_{i} W_{j}b_{j}+\mbox{h.c.})
=\sum_{i,\mu}(b^{\dagger}_{i+\mu}  e^{i\delta A_{i,\mu}}b_{i}+\mbox{h.c.}),
\label{CBhopping}
\end{eqnarray}
where
\begin{eqnarray}
\delta A_{i,\mu}={2\pi \over \nu_{\rm ep}}
\epsilon_{\mu\lambda}\sum_{i'}\nabla_{\lambda}G(r,r')
\biggl(\sqrt{\rho_{\rm ep}}(\eta^{\dagger}_{i'}+\eta_{i'})
+\eta^{\dagger}_{i'}\eta_{i'}\biggr).\nonumber\\
\label{CSA}
\end{eqnarray}
Substituting Eq.(\ref{CBDC}) into the Hamiltonian (\ref{BHCS1}), 
and keeping terms up to the second order of the field $\eta_{i}$,
we obtain the effective Hamiltonian with the quadratic order of $\eta_{i}$,
\begin{eqnarray}
&&H_{{\rm eBHM}}\rightarrow H_{\eta}=-{\tilde J}\sum_{i,\mu}
\biggl[-(\nabla_{\mu}\eta^{\dagger}_{i})(\nabla_{\mu}\eta^{\dagger}_{i})\nonumber\\
&&-i\sqrt{\rho_{\rm ep}}\eta^{\dagger}_{i}(\nabla_{\mu}\delta A_{i,\mu})
+i\sqrt{\rho_{\rm ep}}\eta_{i}(\nabla_{\mu}\delta A_{i,\mu})
-\rho_{\rm ep}(\delta A_{i,\mu})^2\biggr],\nonumber\\
\end{eqnarray}
where we have neglected an irrelevant constant.
By using the properties of the Green function, 
$$(\nabla_{\mu})^{2}G(r,r')=-\delta_{rr'}, \; \mbox{and} \; 
\epsilon_{\mu\lambda }\nabla_{\mu} \nabla_{\lambda}  G(r,r')=0,$$
we obtain the final form,
\begin{eqnarray}
H_{\eta}&=&\sum_{i,\mu}{\tilde J}(\nabla_{\mu}\eta^{\dagger}_{i})(\nabla_{\mu}\eta^{\dagger}_{i})\nonumber\\
&+&\sum_{i,i'}\tilde{J}(2\pi f)^{2}
(\eta^{\dagger}_{i}+\eta_{i})G(r,r')(\eta^{\dagger}_{i'}+\eta_{i'}).\nonumber\\
\label{Heta1}
\end{eqnarray}
The second term of this Hamiltonian $H_\eta$ is the contribution from
the CS gauge field $\delta A_{i,\mu}$.
As we show, this CS gauge coupling gives a finite mass to the `would-be 
massless Nambu-Goldstone boson' as a result of the long-range 
interactions \cite{Ezawa,Ichinose1}. 

We use the Fourier-transformed representation of $H_\eta$.
The lattice Green function $G(r,r')$ is explicitly given as \cite{Kogut, Savit},
\begin{eqnarray}
G(r,r')&=&\int \frac{d^{2}k}{(2\pi)^{2}}
\frac{e^{ik\cdot (r-r')}}{4-2\sum_{\mu}\cos (k\cdot \mu)}.
\label{green}
\end{eqnarray}
It should be noticed that this function has a infrared singularity but its
derivative is well-defined.
Substituting Eq.(\ref{green}) into Eq.(\ref{Heta1}) and then taking  the Fourier transformation, we obtain the following Hamiltonian 
by using the Nambu representation $\vec{\eta}=(\eta(k),\eta^{\dagger}(-k))^t$,
\begin{eqnarray}
&&H_{\eta}=\int_{k>0}\frac{d^{2}k}{(2\pi)^{2}}\vec{ \eta}^{\dagger}(k){\hat H}_{\eta}\vec{\eta}(k).\\
&&{\hat H}_{\eta}\equiv 
\begin{bmatrix}
  \epsilon (k)+2\alpha & 2\alpha \\
  2\alpha & \epsilon (k)+2\alpha
\end{bmatrix},\nonumber\\
&&\epsilon(k)\equiv \tilde{J}\biggl[4-2\sum_{\mu}\cos(k\cdot \mu)\biggr], \label{ep}\nonumber\\
&&\alpha=\frac{\tilde{J}(2\pi f)^{2}}{4-2\sum_{\mu}\cos (k\cdot \mu)}\nonumber.
\label{al}
\end{eqnarray} 
It is easy to carry out the Bogoliubov transformation for 
this matrix ${\hat H}_{\eta}$;
we calculate the eigenvalues of the matrix $\sigma_{3}{\hat H}_{\eta}$
to obtain the excitation energy, 
where the $\sigma_{3}$ is the $z$-component of the Pauli matrix \cite{Altland}. 
This computation preserves the Bose commutation relation. 
Thus the matrix is directly diagonalized by multiplying a unitary operator 
${\hat {\cal U}}$ and the excitation energy is obtained as,
\begin{eqnarray}
&&{\hat {\cal U}}\sigma_{3}{\hat H}{\hat {\cal U}}^{\dagger}=
\begin{bmatrix}
  E(k) & 0 \\
  0 & E (-k)
\end{bmatrix},\nonumber\\
&&E(k)= \biggl[( \epsilon (k)+2\alpha)^{2}-4\alpha^{2}\biggr]^{\frac{1}{2}}.
\label{CBex}
\end{eqnarray}
By taking the long wave limit, we have
\begin{eqnarray}
E(k)\xrightarrow{{\bf k}\rightarrow 0 } 2(n+1)J(2\pi f)=2(n+1)J\Phi, 
\label{CBgap}
\end{eqnarray}
where $\Phi=2\pi f$ is the magnitude of the magnetic flux per plaquette.
The above result indicates that the excitation energy of the excess CB is gapped.
We plot the energy spectrum (\ref{CBex}) in Fig.\ref{CB_excitation} (a). 

Here we should remark that in the vicinity of Mott state there are two 
independent gapped excitations: 
the one comes from this CB boson sector and the other from the based Mott state.
However, the former gap only depends on the parameter $J$ as in Eq.(\ref{CBgap}).
Therefore, the measurement of this energy gap seems feasible in recent experiments. 
This point will be discussed in Sec.VII.

Let us see the effect of the NN repulsion.
It is not difficult to introduce the NN repulsion,  $V\sum_{i,j}n_{i}n_{j}$, 
in the above calculation.
In particular for the case $U>V$, the calculation is rather straightforward.
By using Eqs.(4.12) and (4.13), 
the chemical potential $\tilde{\mu}$ is estimated as
$\tilde{\mu}=-4\tilde{J}+4V\rho_{\rm ep}$.
Then, the quadratic terms of the fluctuation $\eta_{i}$ come from the term $V\sum_{i,j}n_{i}n_{j}$ are obtained as follows,
\begin{eqnarray}
V\biggl[\sum_{i}(4\rho_{\rm ep}\eta^{\dagger}_{i}\eta_{i})+\sum_{\langle i,j\rangle}
\rho_{\rm ep}(\eta^{\dagger}_{i}+\eta_{i})(\eta^{\dagger}_{j}+\eta_{j})\biggr].
\label{add_V}
\end{eqnarray}
Adding these terms to Eq.(\ref{Heta1}) and using the Bogolibov transformation 
as before, we obtain the excitation spectrum including $V$-term as, 
\begin{eqnarray}
E_{V}(k)&=& \biggl[( \epsilon (k)+2\alpha+2\rho_{\rm ep}V
\gamma(k) )^{2}\nonumber\\
&&-(2\alpha+2\rho_{\rm ep}V\gamma(k))^{2}\biggr]^{\frac{1}{2}}.
\nonumber\\
\label{CBexV}
\end{eqnarray}
where $\gamma(k)=\sum_{\mu=1,2}\cos(k\cdot \mu)$. 
In Fig. \ref{CB_excitation} (b), we plot $E_{V}(k)$. 
The value of the energy gaps $E(0)$ and $E_{V}(0)$ are the same, 
but as shown in Fig.\ref{CB_excitation} (c), 
the curvature of $E_{V}(k)$ around ${\bf k}={\bf 0}$ is larger than that of $E(k)$.
Therefore, we expect that the ground-state of a finite NN repulsion system
is more stable than that of $V=0$ 
as excitations with a finite momentum are suppressed by the NN repulsion. 

In Sec.V, based on the study of the CS theory in this section,
we shall introduce a wave function of the CB. 
It has a form of the Gutzwiller type but contains strong correlations between 
bosons as described by the CS gauge theory.
We calculate the ground-state energy of the states and compare it with
that of the obtained states in Sec.III.
Through the comparison, we can judge which state is a better candidate for the 
ground-state.


\section{Chern-Simons Gutzwiller approximation of excess particle} \label{CV}
\setcounter{equation}{0}

In this section, we shall formulate the CS-Gutzwiller theory for the excess
particle system whose Hamiltonian is given by Eq.(\ref{eBH}).
Wave function for the bosonic analog of the FQH state is constructed by
using the `singular gauge transformation' similarly to Eq.(\ref{CSboson}) in Sec.IV.
Then we calculate the energy of the ground-state and compare it with that
of the state obtained by the simple Gutzwiller approximation in Sec.III.
We consider the both $V=0$ and $V=0.2$ cases.
This formulation is nothing but the bosonic counterpart of the CB
approach for the electron FQH state.


\subsection{Chern-Simons transformation and the Gutzwllier approximation}

In the CS theory for the CB, fictitious flux quanta is attached to particle. 
As a result, the CBs have strong correlation with each other
through the Aharanov-Bohm effect.
In the present case, the number of the attached flux quanta is 
$1/\nu_{\rm ep}$.
In the mean-field approximation, the external magnetic field and the magnetic
field of the fictitious gauge field (the CS gauge field) cancel out with each other,
and the homogeneous BEC of the CB is a possible ground-state of the system.

Let us recall the excess particle Hamiltonian $H_{\rm eBHM}$, and the Gutzwiller-wave function 
$|\Psi_{\rm GW}\rangle$,
\begin{eqnarray}
H_{{\rm eBHM}}&=&-J(n+1)\sum_{i} (c^{\dagger}_{i+\mu} e^{iA_{i,\mu}}c_{i}
+{\rm h.c.})\nonumber\\
&&+\sum_{i}U (n_{i}-1)n_{i}+V\sum_{\langle i,j \rangle}n_{c,i}n_{c,j} \nonumber\\
&&-{\tilde \mu} n_{i},  \nonumber  \\
|\Psi_{\rm GW}\rangle &=& \prod_{i=1}^N\biggl( \sum^{n_{c}}_{n=0} 
f^{i}_{n}|n\rangle_{i}\biggr). \nonumber
\end{eqnarray}
The above wave function $|\Psi_{\rm GW}\rangle$ is site-factorized 
and no correlation exists in particles at different sites.
In order to attach the flux quanta to particles, we introduce the following
unitary transformation $U_{\rm G}$,
\begin{eqnarray}
U_{G}&=& \prod_{i=1}^N W_{i},\\
W_{i}&=&e^{i\nu_{\rm ep}^{-1}\sum_{j\neq i}\theta(i,j)n_{j}}, 
\end{eqnarray}
which is the first-quantization representation of the operator $U_i$ in 
Eq.(\ref{CSboson}).
This transformation $U_{G}$ is nothing but the CS transformation 
on the lattice \cite{Fradkin}.
By applying the unitary transformation $U_{\rm G}$ to the simple
Gutzwiller-wave function $|\Psi_{\rm GW}\rangle$, the flux-attached wave function,
$|\Psi_{\rm CS}\rangle$, which we call CS wave function, is produced,
\begin{eqnarray}
|\Psi_{\rm CS}\rangle &=& \prod_{i=1}^N W_{i}|\Psi_{\rm GW}\rangle
= \prod_{i=1}^N\biggl( \sum^{n_{c}}_{n=0} W_{i}f^{i}_{n}|n\rangle_{i}\biggr) \nonumber\\
&\equiv& \prod_{i=1}^N\biggl( \sum^{n_{c}}_{n=0} \gamma^{i}_{n}|n\rangle_{i}\biggr), 
\label{PsiCS}\\
\gamma^{i}_{n}&\equiv& e^{i \nu_{\rm ep}^{-1}\sum_{j\neq i}\theta(i,j)n_{j}}f^{i}_{n}.
\label{gamma}  
\end{eqnarray}
As $\{\gamma^i_n\}$ in Eq.(\ref{gamma}) show, $|\Psi_{\rm CS}\rangle$ represents
a strongly correlated state.
See Fig.\ref{CStheory}.
The state $|\Psi_{\rm CS}\rangle$ is a candidate for the ground-state
for specific fillings, and physical quantities like energy are calculated as 
\begin{equation}
E_{\rm CS}=\langle \Psi_{\rm CS}|H_{\rm eBHM}|\Psi_{\rm CS}\rangle.
\label{ECS}
\end{equation}
In the practical calculation, we employ the periodic boundary condition.
The summation in Eq.(\ref{gamma}) takes only once for each site $j\neq i$.

In the following subsection, we obtain the CS wave function
of the ground-state by the Gutzwllier method and calculate its energy.

\subsection{Numerical results}

We have two candidates for the ground-state within the Gutzwllier method,
one is $|\Psi_{\rm GW}\rangle$ and the other is  $|\Psi_{\rm CS}\rangle$,
which are defined by Eqs.(\ref{Gwf}) and (\ref{PsiCS}), respectively.
We calculated the energy of the two states by the Gutzwllier approximation. 
To see which state has a lower energy, we have to carefully define 
the energy of excess particles.

For the state $|\Psi_{\rm GW}\rangle$, the energy of the MI has to be subtracted.
In the MI, the particle number at each site is unity with only very small fluctuations.
Also the local density terms including the chemical potential term, 
$-({U \over 2}+\mu)n_i$, should be subtracted from the total energy as they 
only contribute to control the average particle density.
Therefore, we define the excess particle energy in the $|\Psi_{\rm GW}\rangle$, 
$E_{\rm ex, GW}$, as 
$$
E_{\rm ex, GW}\equiv \langle H_{\rm BHM}\rangle 
+(\mu+U/2)\sum_{i}{\rho}-\frac{U}{2}(1-1)\cdot 1
- V\sum_{\langle i,j\rangle}1\cdot 1,
$$
where ${\rho}=1.25$ in the present case.
On the other hand for the state $|\Psi_{\rm CS}\rangle$,  
$$
E_{\rm ex, CS}\equiv E_{\rm CS}+(\tilde{\mu}+U)\rho_{\rm ep}.
$$

As we stated above, we consider the two cases $V/U=0$ and $V/U=0.2$.
If the particle density were sufficiently large, the vortex-lattice states observed
for $V=0$ in the magnetic field $f=1/2$ and $1/3$ were expected to be the
genuine ground-state, i.e., the optical lattice plays a role of the vortex pinning
and stabilizes the vortex solid.
However in the present system, the density of the excess particle is very low, 
and therefore it is interesting to compare the vortex-lattice state in Sec.III to 
the CS ground-state.
On the other hand for the case with $V/U=0.2$, we expect that the CS state
$|\Psi_{\rm CS}\rangle$ has a lower energy than $|\Psi_{\rm GW}\rangle$
as there exists no order in the state $|\Psi_{\rm GW}\rangle$.

\begin{figure}[t]
\centering
\includegraphics[width=8cm]{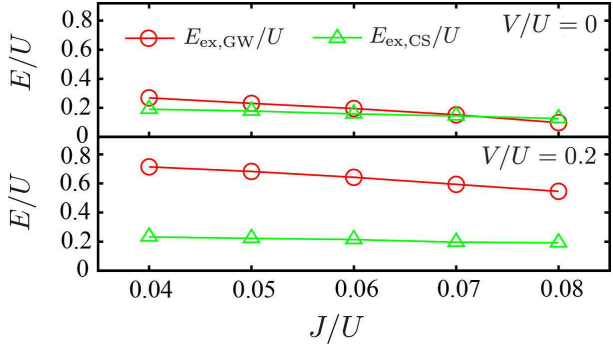}
\caption{(Color online)
Energies of the states described by the Gutzwiller-wave function and the CS wave
function, $E_{\rm ex, GW}$ and $E_{\rm ex, CS}$, respectively.
For $V=0$, these two states have comparable energy.
On the other hand for $V/U=0.2$, the CB state has a lower energy than the
GW-function state..
}
\label{comparison}
\end{figure}

The numerical result for $0.04\leq J/U \leq 0.08$ is shown in Fig.\ref{comparison}. 
For $V=0$ case, we find that both energies $E_{\rm ex,GW}$ and $E_{\rm ex,CS}$ are very close, i.e.,
the vortex solid phase competes with the excess particle FQH state. 
On the other hand for the $V/U=0.2$ case, the energy of the state 
$|\Psi_{\rm CS}\rangle$, $E_{\rm ex, CS}$, is lower than $E_{\rm ex,GW}$ of 
the Bose-metal phase. 
{\em This result indicates that the finite NN repulsion prefers the 
bosonic analogs of the FQH state.} 
This is one of the main conclusion of the present paper.

\section{Composite fermion picture}\label{CFpicture} \label{CVI}
\setcounter{equation}{0}

In this section, we continue the analytically study on the excess particle BHM with a relatively 
large NN repulsion $V$.
As we showed in Sec.IV, the non-SF phase numerically observed 
in Sec.III for $V/U=0.2$ is not a true ground-state, and instead of it,
the strongly-correlated state, which is described by the CS wave function,
is a good candidate for the genuine ground-state. 

The above study is based on the CB picture described by the CS gauge theory
coupled with bosons.
In this section, we employ the CF picture, which is 
another possible theory describing the FQH state of the hard-core bosons.
In fact, the exact diagonalization for the system with a small size exhibits
a good overlap between the CF wave function and that of the excess particle BHM  \cite{Cooper2}.
In this section, we shall explain how the CF picture appears from the effective
Hamiltonian of the excess particle in Eq.(\ref{eBH}).
In the previous paper \cite{IchinoseMatsui}, 
we studied dynamics of electrons in the half-filled Landau
level, and showed that the CF picture appears as a result of `the particle-flux
separation', which is a similar phenomenon to the spin-charge separation in 
the strongly-correlated systems like the high-$T_c$ cuprates.

We introduce a {\em fermion} $\psi_i$ that is defined as follows,
\begin{eqnarray}
\psi_i&=&\tilde{W}_ic_i,  \;\; c_i=\tilde{W}_i^\dagger \psi_i,   \nonumber \\
\tilde{W}_i&=&\exp\Big[ip\sum_{r'}\theta(i,r')c^\dagger_{i'}c_{i'}\Big]
\nonumber \\
&=&\exp\Big[ip\sum_{r'}\theta(i,r')\psi^\dagger_{i'}\psi_{i'}\Big],
\label{CFop}
\end{eqnarray}
where $p$ is {\em an odd integer} that is determined shortly and the other notations
are the same with those in Sec.IV.
Then it is not so difficult to show that $\psi_i$'s satisfy the fermionic 
anti-commutation relations, and the original boson is expressed 
as a composite of $\psi_i$ and the $p$-flux quanta.
One may think that the strong on-site repulsion generating the MI
produces a fermionic properties of the excess particles, but the on-site
repulsion itself is not enough to generate the CF picture as we see in this section.
The hopping term of the Hamiltonian, $H_{\rm J}$, is expressed as follows in terms
of $\psi_i$,
\begin{eqnarray}
H_{\rm J}&=&-\tilde{J}\sum_{i,\mu}(\psi^\dagger_i\tilde{W}_i\tilde{W}_{i+\mu}^\dagger)
e^{iA^{\rm ex}_{i,\mu}}\psi_{i+\mu}+\mbox{h.c.}  \nonumber \\
&=&-\tilde{J}\sum_{i,\mu}(\psi^\dagger_i e^{-iA^{{\rm eff}}_{i,\mu}}\psi_{i+\mu},
+\mbox{h.c.}),  \label{CF1}\\
A^{\rm ex}_{i,\mu}&=&\sum_{i'}\nabla_\mu\theta(i,r')f, \;\;
{\rm rot} \ A^{\rm ex}_{i,\mu}=2\pi f, \nonumber \\
A^{{\rm eff}}_{i,\mu}&=& \sum_{i'} \nabla_{\mu}\theta(i,r') 
\biggl[p\psi^{\dagger}_{i'}\psi_{i'}-f\biggr], 
\label{CFH}
\end{eqnarray}
where we have used Eq.(\ref{angleG}) and 
$A^{\rm ex}$ denotes the vector potential of the external magnetic field 
in the symmetric gauge.
From Eq.(\ref{CFH}), it is obvious that when the fermion $\psi_i$ has a
homogeneous distribution with the average density per site $\rho_{\rm ep}$ and also
the parameters satisfy the relation $f=(p-1)\rho_{\rm ep}$, we have
$\langle {\rm rot} \ A^{\rm eff}_{i,\mu}\rangle =2\pi \rho_{\rm ep}$. 
Therefore, {\em the fermions $\psi_i$ fill just the Hofstadter bands ramifying
from the lowest Landau level
if interactions between $\psi_i$ is irrelevant}.
However, the fermion $\psi_i$ has a nonlocal interaction with each other
through $A^{{\rm eff}}_{i,\mu}$ in Eq.(\ref{CFH}), and therefore an elaborate
discussion is needed to justify the above assumption.

To study the above strongly-correlated fermion system, we introduce
the following slave-particle representation, 
\begin{equation}
\psi_i=\phi_i\zeta_i,
\label{slave}
\end{equation}
where $\phi_i$ is a hard-core boson and $\zeta_i$ is a fermion, and 
we call $\zeta_i$ and $\phi_i$ chargon and fluxon, respectively.
It is not so difficult to show that $\psi_i$'s in Eq.(\ref{slave}) satisfy the 
fermionic anti-commutation relation.
Physical state condition of the slave-particle Hilbert space is given 
by the local constraint, 
\begin{equation}
\zeta^\dagger_i \zeta_i=\phi^\dagger_i \phi_i.
\label{const}
\end{equation}
By the local constraint Eq.(\ref{const}), we can prove
\begin{equation}
\psi^\dagger_i\psi_i=\zeta^\dagger_i\zeta_i\phi^\dagger_i\phi_i
=\zeta^\dagger_i\zeta_i\zeta^\dagger_i\zeta_i=\zeta^\dagger_i\zeta_i
=\phi^\dagger_i\phi_i.
\label{densityeq}
\end{equation}
Then the nonlocal operator $\tilde{W}_i$ in Eq.(\ref{CFop}) is expressed as 
\begin{equation}
\tilde{W}_i=\exp\Big[ip\sum_{r'}\theta(i,r')\phi^\dagger_{i'}\phi_{i'}\Big]
\equiv W^\phi_i.
\label{Wx2}
\end{equation}
The Hamiltonian $H_{\rm eBHM}$ is expressed as follows in the slave-particle representation,
\begin{eqnarray}
H_{\zeta\phi}&=&-J\sum(\zeta^\dagger_{i+\mu}\phi^\dagger_{i+\mu}W^\phi_{i+\mu}
W^{\phi\dagger}_{i}e^{iA^{\rm ex}} \phi_i\zeta_i+\mbox{h.c.})  \nonumber \\
&&-L_{\rm int} 
-\sum(\mu_\zeta \zeta^\dagger_i\zeta_i+\mu_\phi\phi^\dagger_i\phi_i) 
\nonumber \\
&&-\sum\lambda_i(\zeta^\dagger_i\zeta_i-\phi^\dagger_i\phi_i).
\label{Hslave}
\end{eqnarray}
where $\lambda_i$ is the Lagrange multiplyer for the local constraint 
Eq.(\ref{const}), and $L_{\rm int}$ denotes the NN repulsion in $H_{\rm eBHM}$
in Eq.(\ref{eBH}).

In order to study the above fermion system, we employ a Lagrangian 
formalism with an imaginary time $\tau$.
The partition function $Z$ and the Lagrangian $L_{\zeta\phi}$ are given as follows,
\begin{eqnarray}
&&Z=\int[d\zeta][d\phi]\exp \Big(\int^\beta_0d\tau L_{\zeta\phi}\Big), 
\nonumber \\
&&L_{\zeta\phi}=-\sum\zeta^\dagger_x \partial_\tau\zeta_x
-\sum\phi^\dagger_x \partial_\tau\phi_x -H_{\zeta\phi},
\label{Zslave}
\end{eqnarray}
where $x$ denotes the 3D coordinate $x=(\tau, i)$ ($\tau\in [0,\beta]$), and
$\beta=1/(k_{\rm B}T)$ with the Boltzmann constant $k_{\rm B}$
and temperature $T$.
Then we apply the following Hubbard-Stratonovich transformation to the above
system $L_{\zeta\phi}$,
\begin{equation}
Z=\int[d\zeta][d\phi][dV]\exp \Big(\int^\beta_0d\tau L_{\zeta\phi V}\Big),
\label{Zslave2}
\end{equation}
where
\begin{eqnarray}
&&L_{\zeta\phi V}=-\sum\zeta^\dagger_x (\partial_\tau+i\lambda_x-\mu_\zeta)\zeta_x
\nonumber \\
&&\hspace{0.5cm}
-\sum\phi^\dagger_x (\partial_\tau-i\lambda_x-\mu_\phi)\phi_x \nonumber \\
&&\hspace{0.5cm}
+J\sum\Big[V_{x \mu}(\phi_{x+\mu}W'_{x+\mu}{W'}^{\dagger}_x\phi^\dagger_x
+\zeta^\dagger_{x+\mu}e^{ia_\mu}\zeta_x)+\mbox{h.c.}\Big]
\nonumber \\
&&\hspace{0.5cm}
+J\sum\Big(\phi^\dagger_{x+\mu}\phi_{x+\mu}\phi^\dagger_x\phi_x+
\zeta^\dagger_{x+\mu}\zeta_{x+\mu}\zeta^\dagger_x\zeta_x\Big)
\nonumber  \\
&&\hspace{0.5cm}
-J\sum|V_{i\mu}|^2+L_{\rm int},
\label{LetaphiV}
\end{eqnarray}
and
\begin{eqnarray}
W'_x&=&W^\phi_x e^{-ip\sum_{r'}\theta(x,r')\rho_{\rm ep}}  \nonumber \\
&=&\exp\Big[ip\sum_{r'}\theta(x,r')(\phi^\dagger_{x'}\phi_{x'}-\rho_{\rm ep})\Big], 
\nonumber \\
a_\mu&=&\sum_{r'}\nabla_\mu\theta(x,r')(p\rho_{\rm ep}-f).
\end{eqnarray}
Several comments on the system $L_{\zeta\phi V}$ in Eq.(\ref{LetaphiV})
are in order.
\begin{enumerate}
\item The fields $\lambda_i$ and $V_{i\mu}$ behaves like a gauge field. 
In fact, $L_{\zeta\phi V}$ {\em is invariant under a time-dependent local gauge
transformation},
\begin{eqnarray}
&&(\zeta_i,\phi_i,V_{i\mu}, \lambda_i) \rightarrow \nonumber \\
&&\hspace{1cm}(e^{i\alpha_i}\zeta_i,e^{-i\alpha_i}\phi_i,
e^{i\nabla_\mu\alpha_i}V_{i\mu}, \lambda_i-\partial_\tau\alpha_i).\nonumber
\end{eqnarray}
\item Low-energy properties of the system is determined by the dynamics
of the gauge field $V_{i\mu}$.
If its dynamic is realized in a deconfinement phase like the Coulomb phase,
the fields $\zeta_i$ and $\phi_i$, chargon and fluxon, describe quasi-excitations,
whereas in the confinement phase, the original boson is the only physically
observable object. 
We call the phenomenon in the former case {\em particle-flux separation}.
\item There appear the NN attractive force in the channel 
($\zeta_{i+\mu}-\zeta_i)$ and $(\phi_{i+\mu}-\phi_i)$.
This attractive force makes the system unstable into a phase separated state
if the particle-flux separation takes place.
In order to make the system stable, {\em the existence of the NN repulsion,
$L_{\rm int}$, is needed}.
\item In the particle-flux separated state, the fluxon $\phi$ is nothing but a fermion
in the commensurate external magnetic field.
This fermion is defined as $\varphi_i=W^\phi_i \phi_i$ and $\varphi_i$ feels 
the effective magnetic field $f_{\varphi}=p\rho_{\rm ep}$.
As the density of $\varphi_i$ is $\rho_{\rm ep}$, $\varphi_i$ fills the $1/p$-levels
in the lowest-Hofstadter bands ramifying from the lowest Landau level 
in the continuum.
\item $a_\mu$ is the vector potential that represents the magnetic field with
flux quanta $(f-p\rho_{\rm ep})=(f-f_{\varphi})$ per plaquette.
Then it is obvious that $\eta_i$ is nothing but the CF if the particle-flux
separation is realized. 
The Hofstadter butterfly \cite{Hofstadter} predicts the parameters
 $(\rho_{\rm ep}, f)$ at which gapful states appear.
\end{enumerate}

The above gauge-theoretical consideration gives a basis of the CF picture
propose by M\"{o}ller and Cooper for the low-filling bosons on the lattice
\cite{Cooper2}.
In their work, $p=1$ and the trial CF-state wave function is given as 
(in their notation)
\begin{equation}
\Psi_{\rm trial}(\{ \vec{r}_i \})
=\Psi_{\rm J}(\{ \vec{r}_i \})\times \Psi_{\rm CF}(\{ \vec{r}_i \}),
\label{MC}
\end{equation}
where both of $\Psi_{\rm J}$ and $\Psi_{\rm CF}$ are fermionic wave functions.
$\Psi_{\rm J}$ comes from the flux attachment and $\Psi_{\rm CF}$ is 
the wave function of the CF.
In our derivation, {\em $\Psi_{\rm J}$ is nothing but the wave function of
$\varphi_i$ and $\Psi_{\rm CF}$ is that of $\zeta_i$}.
{\em Our study in this section has clarified the condition that the CF 
picture appears as quasi-excitations at low energy}.
Problem of the gauge dynamics of the system $L_{\zeta\phi V}$ can be studied
by the hopping expansion and the realization of the deconfinement phase
(Coulomb like phase) is suggested at low temperature \cite{Ichinose2}.
However, a more detailed study is needed to reach a decisive conclusion.
This problem is under study and the results will be published in near future.

There is an ambiguity in the way of decoupling the hopping term in $H_{\zeta\phi}$
in Eq.(\ref{Hslave}) by the auxiliary field $V_{i\mu}$.
We call the decoupling in Eq.(\ref{LetaphiV}) {\em optimal particle-flux separation}.
In fact in $L_{\zeta\phi V}$ in Eq.(\ref{LetaphiV}), the chargon and fluxon 
do not interact with each other except the gauge interaction through $V_{i\mu}$.
In the previous paper, we studied the {\em half-filled Landau level state} of 2D
electron systems in a strong magnetic field.
There we used another decoupling in which the chargon feels magnetic fluxes
carried by the fluxon.
As a result of a BEC of the fluxon, the external magnetic field is totally
shielded and the CF behaves like a gapless fermion with a Fermi line in the
momentum space.
For this case, it is known that the dynamics of the gauge field $V_{i\mu}$ realizes
a deconfinement phase and therefore the CF picture is justified \cite{Ichinose2}.

At present, relationship between the CB and CF approaches is not clear.
We shall study this problem in detail and hope that experiments on the cold
atomic gases give an important clue to solve this problem.

\section{Discussion and Conclusion} \label{CVII}
\setcounter{equation}{0}

In this paper we have studied the ground-state properties of lattice bosons 
in the strong magnetic field in the vicinity of the Mott states.
We have first rederived the excess particle effective Hamiltonian from the 
BHM with the NN repulsions. 
By using the Gutzwiller numerical method, we obtained the phase diagrams and
investigated the ground-state properties for particular points near the Mott lobes 
for which the appearance of the bosonic analogs of the FQH state is expected.
We have found that the vortex solids form in the absence of the NN repulsion, 
but a finite NN repulsion destabilizes the vortex solids and the featureless 
homogeneous state appears as the ground-state of the Gutzwiller-wave function
that we call the Bose-metal. 

In order to investigate the ground-state in the system with finite NN repulsions
in detail,
we have made use of the CS theory for the excess particle system.
After the analytical study of the CS theory for the lattice boson in the strong 
magnetic field, we have applied the CS theory to the Gutzwiller numerical method
and proposed the CS wave function for describing the bosonic FQH state.
Then we calculated the ground-state energies of the state given by the Gutzwiller-wave function
and the state of the CS wave function.
We found that the NN repulsions prefer the state of the CS wave function.

We expect that the measurement of the energy gaps calculated in 
Sec.\ref{CIII}, \ref{CIV} and \ref{CVI} is feasible in real experiments on 
ultra-cold atomic gases. 
As one example, the lattice modulation method inducing 
the two photon Bragg-spectroscopy \cite{Stoferle,Higgs} may be efficient.
If the bosonic FQH state or the Bose-metal forms in real experiments, 
the total system has two energy gaps, i.e., one is 
the particle-hole excitation gap $U$ in the base Mott state and
the other is the excitation gap of the excess particle. 
On the other hand, if the system forms the vortex solid, i.e.,
the superfluid of the finite-momentum mode BEC, there exists
a gapless excitation as a result of the spontaneous breaking of the U(1) symmetry. 

Finally, we have studied the CF theory for the excess particle by using the CS theory.
Our previous gauge-theoretical study on the electron system of the half-filled Landau
level \cite{IchinoseMatsui} is applicable rather straightforwardly to the present
boson systems, and we showed the condition that the CF appears as quasi-excitations 
at low temperature.

We shall study the BHM with the NN repulsions by means of the exact diagonalization, 
the cluster Gutwiller method, etc., and examine the obtained results in this paper.
In particular at present, the relationship between the CB and CF approaches to the 2D
strongly-correlated systems in a strong magnetic field is not understood.
We expect that the ultra-cold atomic system plays an important role 
to solve this problem because of its controllability and  versatility.
We shall study this problem by means of the analytical and numerical methods
mentioned above and propose experimental set ups for testing the CB and CF
pictures.

\appendix
\renewcommand{\thefigure}{\Alph{section}.\arabic{figure}}
\setcounter{figure}{0}
\renewcommand{\theequation}{A.\arabic{equation}}
\section{Estimation of parameters of BHM for real experiments}
In this appendix, we microscopically evaluate the on-site and nearest-neighbor (NN)
interactions (i.e., $U$ and $V$) in the BHM of Eq.(\ref{BH})
as the ratio $V/U$ plays an important role in the present work for experimental
realization. 
We assume the ratio $V/U\sim 0.2$, and 
therefore it is needed that the NN interaction $V$ is comparable with the on-site 
interaction $U$.
As we show, in real experiment, such a condition is feasible 
by using large magnetic dipolar atoms like Cr, Er, and Dy \cite{Lahaye}.
In fact, by using dipolar atoms and the Feshbach resonance techniques, one can 
control the ratio $V/U$ rather freely.
Generally in a dipolar atom system, the on-site $U$ is given as $U=U_{\rm s}+U_{\rm d}$, 
where $U_{\rm s}$ is the contribution from the s-wave scattering and $U_{\rm d}$ 
is the contribution from the dipole-dipole interaction. 
Also the dipole-dipole interaction gives the NN interaction in the lattice system. 
In the case that the dipoles of atoms are perpendicular to the two dimensional plane,
the NN interaction is given as followings,
\begin{eqnarray}
&&V=\int d{\bf r}d{\bf r}' |w_{i}({\bf r})|^{2}
\biggl[\frac{\mu_{0}\mu^{2}G_{{\bf r}-{\bf r}'}}{4\pi |{\bf r}-{\bf r}'|^{3}}\biggr] |
w_{j}({\bf r}')|^{2},  \label{micro_V} \\
&&G_{{\bf r}-{\bf r}'}=1-3\cos^2\theta_{{\bf r}-{\bf r}'},
\label{micro_V2}
\end{eqnarray}
where 
$\mu_{0}$ and $\mu$ are the permeability of vacuum and magnetic permeability of 
the dipolar atom, respectively, and $\theta_{{\bf r}-{\bf r}'}$ is the angle
between $({\bf r}-{\bf r}')$ and the orientation of the dipole.
$w_{i(j)}({\bf r})$ is the lowest-band Wannier function, which is tightly localized 
at site $i (j)$. 
As the above overlap integral (\ref{micro_V}) shows, the value $V$ is determined by the choice of dipolar atom.

\begin{figure}[t]
\centering
\includegraphics[width=7.7cm]{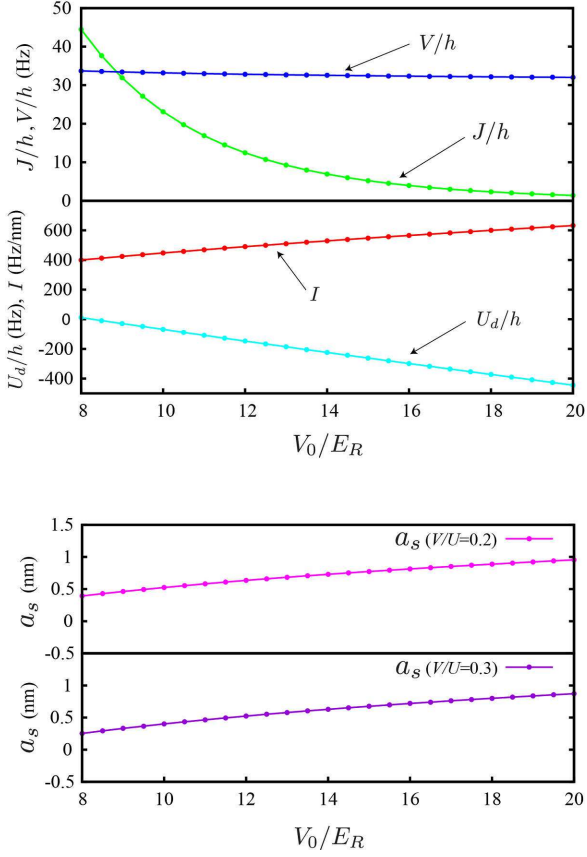}
\caption{Parameters of BHM evaluated microscopically as a function of $V_0/E_{\rm R}$
and the s-wave scattering length $a_{s}$ for the experimental setup. 
The value of $a_s$ is obtained as $a_{s}(V_{0})=[V/\alpha-U_{d}]/I$, 
where $\alpha=V/U=0.2$ or $0.3$.}
\label{BHMparameter}
\end{figure}

In order to increase the ratio $V/U=V/(U_{\rm s}+U_{\rm d})$, it is necessary 
to reduce the value of $U_{\rm s}+U_{\rm d}$. 
This is feasible in real experiments by controlling the parameter $U_{\rm s}$, i.e.,
by controlling the s-wave scattering length $a_s$ by the Feshbach
resonance techniques \cite{optical,Inouye}.  
Even if the value $U_{\rm d}$ has a large positive value compared to $V$, the small or negative value of $U_{\rm s}$ 
can reduce the total value of $U_{\rm s}+U_{\rm d}$.
As we show, $U_{\rm d}$ has a strong dependence on the optical lattice potential
and it can have even a negative value.

For future experiments, we shall estimate the BHM parameters, $J$, $U$, and $V$. 
In particular, we estimate the value of tunable s-wave scattering length $a_{s}$ for $V/U=0.2$ and $0.3$ to be realized. 

Then, we consider a two dimensional lattice similar to the recent experimental 
setup \cite{Baier}, and
${}^{168}$Er for the dipolar atom with the moment $\mu=7\mu_{\rm B}$ 
($\mu_{\rm B}$ is Bohr magneton).
We consider two dimensional optical lattice potential with the lattice spacing 
$d=266$ [nm]. 
This potential is explicitly given as
\begin{equation}
V({\bf r})=V_{0}[\cos^{2}((2\pi/\lambda)x) + \cos^{2}((2\pi/\lambda)y)]
+\frac{1}{2}m\omega^{2}_{z}z^{2},
\end{equation}
where $V_{0}$ is the potential depth 
and $\lambda$ is the laser wave length, $m$ is the atom mass, and $\omega_{z}$ is 
the frequency of the harmonic trap used to construct 
quasi-two dimensional system. 
Here, the optical lattice spacing is given as $d=\lambda/2$. 
In this system, the other BHM parameters are given by the overlap integrals similar
to Eq.(\ref{micro_V}) \cite{Baier}
\begin{eqnarray}
J&=& -\int d{\bf r} w^{*}_{i}({\bf r})\biggl[-\frac{{\hbar}^{2}\nabla^{2}}{2m}+V({\bf r}) \biggr] w_{j}({\bf r}),
\label{micro_J}\\
U_{s}&=& \frac{4\pi\hbar^{2}a_{s}}{m} \int d{\bf r} |w_{i}({\bf r})|^{4}\equiv a_{s}I,
\label{micro_Us}\\
U_{\rm d}&=&\int d{\bf r}d{\bf r}' |w_{i}({\bf r})|^{2}
\biggl[\frac{49\mu_{0}\mu^{2}_{B}}{4\pi |{\bf r}-{\bf r}'|^{3}}G_{{\bf r}-{\bf r}'}\biggr]
 |w_{i}({\bf r}')|^{2}, \nonumber \\
\label{micro_V3}
\end{eqnarray}
where $m=2.78 \times 10^{-25}$ [kg] is the  ${}^{168}$Er atom mass.
To estimate the above integrals, we employ the harmonic oscillator approximation 
for the optical lattice potential $V({\bf r})$.
In this approximation, the Wannier function $w_{i}({\bf r})$ is replaced by
the harmonic-oscillator wave function of the lowest energy, 
\begin{equation}
w_{i}({\bf r})=\sqrt{\frac{\beta}{\pi}}e^{-\frac{\beta}{2}((x-x_{i})^{2}+(y-y_{i})^{2})}
\times
\biggl[\frac{\beta_{z}}{\pi}\biggr]^{1/4}e^{-\frac{\beta_{z}}{2}(z-z_{i})^{2}},
\label{wani}
\end{equation}
where $\beta\equiv \frac{2m}{\hbar^{2}}\sqrt{E_{\rm R}V_{0}}$, 
($E_{\rm R}$ is the recoil energy $\equiv h^{2}/(2m\lambda^{2})\sim h\times 4.2$[kHz] 
for ${}^{168}$Er ), 
$\beta_{z}\equiv \frac{m\omega_{z}}{\hbar}$ and the spatial coordinate $(x_{i},y_{i},z_{i})$ 
is the three-dimensional coordinate of optical lattice site $i$.  
We take $\omega_{z}\sim 160$ [kHz] to confine atoms tightly in
the two-dimensional plane \cite{Dutta}. 

We numerically calculated the above integrals Eq.(\ref{micro_V}) and 
Eqs.(\ref{micro_J})-(\ref{micro_V3}), and estimated the value $a_{s}$ 
for the ratio $V/U=0.2$ and $0.3$.
We verified that our estimation of the parameters is in good agreement with the 
previous works \cite{Dutta}.
Figure~\ref{BHMparameter} shows the calculated results of the values of $J, V, I$
and $U_{\rm d}$ as a function of the potential depth $V_0$.
$U_{\rm d}$ is negative as $V_0/E_{\rm R}$ is getting large as mentioned above.
The values of $a_{s}$ for realizing the ration $V/U=0.2$ and $0.3$ are also shown there.
From the results, to achieve $J/U=0.05$ and the target ratio $V/U$, 
we found that for $V/U=0.2$,
$a_{s}\sim$0.70[nm] and $V_{0}\sim13.5E_{\rm R}$, and for $V/U=0.3$,
$a_{s}\sim$0.67[nm] and $V_{0}\sim15.0E_{\rm R}$. 
These values of $a_{s}$ and $V_{0}$ are feasible for real experiments.

\acknowledgments
Y. K. acknowledges the support of a Grant-in-Aid for JSPS
Fellows (No.15J07370).
This work was partially supported by JSPS KAKENHI
Grant Number JP26400246.



\begin{references}
\bibitem{TWA0} This concern also appears in solving the Gross-Pitaevskii (GP) equation. 
In the GP community, to avoid this concern an extended method was invented: 
The Truncated-Wigner approximation \cite{TWA}.
\end{references}
\end{document}